 \documentclass[preprint,aps]{revtex4}
 
\begin{document}
 \newcommand{\Qed}{\rule{2.5mm}{3mm}}
 \newcommand{\balpha}{\mbox{\boldmath {$\alpha$}}}
 \def\Tr{{\rm Tr}}
 \def\(#1)#2{{\stackrel{#2}{(#1)}}}
 \def\[#1]#2{{\stackrel{#2}{[#1]}}}
 \def\A{{\cal A}}
 \def\B{{\cal B}}
 \def\Sb#1{_{\lower 1.5pt \hbox{$\scriptstyle#1$}}}
 \draft 
\title{Particular boundary condition ensures that a  fermion in $d=1+5$, 
compactified on a finite  disk, manifests in $d=1+3$ as massless spinor with a charge $1/2$,  
 mass protected and chirally coupled to the gauge field. 
}
\author{ N.S. Manko\v c Bor\v stnik}
\address{ Department of Physics, FMF, University of
Ljubljana, Jadranska 19,Ljubljana, 1000}
\author{ H. B. Nielsen}
\address{Department of Physics, Niels Bohr Institute,
Blegdamsvej 17,\\
Copenhagen, DK-2100}



\begin{abstract} 
The genuine Kaluza-Klein-like theories---with no fields in addition to gravity---have difficulties 
with the existence of massless spinors after the compactification 
of some space dimensions \cite{witten}. We proposed in ref.~\cite{hnkk06}  a  
boundary condition for spinors in $(1+5)$ compactified on a flat disk that ensures  
masslessness of  spinors (with all positive half integer charges) in $d=(1+3)$ as well as their   
chiral coupling to the corresponding background gauge gravitational field.   
In this paper we study the same toy model, proposing a boundary condition allowing a 
massless spinor of  one handedness and only one charge ($1/2$) and  
infinitely many massive spinors of the same charge, allowing disc to be curved.   
We define the 
operator of momentum to be  Hermitean on the vector space of spinor states---the solutions 
on a disc with the boundary. 
\end{abstract}

\maketitle



\date{\today}

\section{Introduction}
\label{introduction}

The major problem of the compactification procedure in all Kaluza-Klein-like theories with 
only gravity and no additional gauge fields is how to ensure that massless spinors 
be mass protected after the compactification. Namely, even if we start with 
only one Weyl spinor in some even dimensional space of  $d=2$ 
modulo $4$ dimensions (i.e.\ in $d=2(2n+1),$ $n=0,1,2,\cdots$) so that there appear no Majorana 
mass if no conserved 
charges exist and families are allowed, as we have proven in 
ref.~\cite{hnm06}, and accordingly with the mass protection from the very beginning, 
a compactification of $m$ dimensions gives rise to a spinor of one handedness in $d$  
with both handedness in $(d-m)$ and is accordingly  not mass protected any longer. 

And  besides, since the spin (or the total conserved angular momentum) 
in the compactified part of space will    
in space of $(d-m)$ dimensions manifest accordingly a charge of both signs while    
in the second quantization procedure antiparticles 
of opposite charges appear anyhow, doubling the number of massless 
spinors when coming from 
$d(=2(2n+1))$-dimensional space down to $d=4$ (after 
the second quantized procedure) is not in agreement  
with what we observe. Therefore there must be some requirements, some boundary 
conditions~\cite{hnkk06}, which 
ensure in a compactification procedure that only spinors of one handedness and one charge 
survive if Kaluza-Klein-like theories have some meaning, what due to the beautifulness of 
the idea of the  gravity as the only  gauge field we would hope for.

One of us~\cite{norma92,norma93,norma94,norma95,Portoroz03,pikanorma06} has for long tried to 
unify the spin and all the charges to only the spin, so that spinors would in 
$d\ge 4$ carry nothing but (two kinds of~\footnote{To understand the appearance of the two kinds 
of spin we invite the reader to look at the 
refs.~\cite{pikanorma06,holgernorma2002,holgernorma2003}.})
a spin and interact accordingly with only the gauge fields of the corresponding 
generators of the infinitesimal transformations (of translations and two kinds of the 
Lorentz transformations in the space of spinors), that is with 
vielbeins $f^{\alpha}{\!}_{a}$ \footnote{$f^{\alpha}{}_{a}$ are inverted vielbeins to 
$e^{a}{}_{\alpha}$ with the properties $e^a{}_{\alpha} f^{\alpha}{\!}_b = \delta^a{\!}_b,\; 
e^a{\!}_{\alpha} f^{\beta}{\!}_a = \delta^{\beta}_{\alpha} $. 
Latin indices  
$a,b,..,m,n,..,s,t,..$ denote a tangent space (a flat index),
while Greek indices $\alpha, \beta,..,\mu, \nu,.. \sigma,\tau ..$ denote an Einstein 
index (a curved index). Letters  from the beginning of both the alphabets
indicate a general index ($a,b,c,..$   and $\alpha, \beta, \gamma,.. $ ), 
from the middle of both the alphabets   
the observed dimensions $0,1,2,3$ ($m,n,..$ and $\mu,\nu,..$), indices from 
the bottom of the alphabets
indicate the compactified dimensions ($s,t,..$ and $\sigma,\tau,..$). 
We assume the signature $\eta^{ab} =
diag\{1,-1,-1,\cdots,-1\}$.
} and  (two kinds of)  
spin connections ($\omega_{ab\alpha}$, which are the gauge fields of 
$S^{ab}= \frac{i}{4}(\gamma^a \gamma^b - \gamma^b \gamma^a)$ and $\tilde{\omega}_{ab\alpha}$, 
which are the gauge fields of 
$\tilde{S}^{ab}= \frac{i}{4}(\tilde{\gamma}^a \tilde{\gamma}^b - \tilde{\gamma}^b \tilde{\gamma}^a), 
$ with $\{\gamma^a, \tilde{\gamma}^b\}_{+}=0$).

In this paper  we take (as we did in ref.~\cite{hnkk06}) for the covariant momentum of a spinor  
\begin{eqnarray}
p_{0 a} &=& f^{\alpha}{\!}_{a}p_{0 \alpha}, \quad p_{0 \alpha} \psi = p_{ \alpha} - \frac{1}{2} S^{cd} 
\omega_{cd \alpha}. 
\label{covp}
\end{eqnarray}
The corresponding Lagrange density ${\cal L}$  for   a Weyl spinor has the form
${\cal L} = E \frac{1}{2} [(\psi^{\dagger}\gamma^0 \gamma^a p_{0a} \psi) + 
(\psi^{\dagger} \gamma^0\gamma^a p_{0 a}
\psi)^{\dagger}]$ and leads to
\begin{eqnarray}
{\cal L} &=& \psi^{\dagger}  \gamma^0 \gamma^a \{E   ( p_{a} - \frac{1}{2} S^{cd}  \omega_{cda}) 
+\frac{1}{2} \{p_{\alpha}, E f^{\alpha}{\!}_a\}_- \}\psi,
\label{weylL}
\end{eqnarray}
with $ E = \det(e^a{\!}_{\alpha}). $ 
If we have no gravity in $d=(1+3)$, and for $\omega_{abc}=0,$ while 
$f^{\sigma}{\!}_s= \delta^{\sigma}_{s} f(\rho)$, 
the equations of motion follow
\begin{eqnarray}
\{E\gamma^0 \gamma^m p_m + E f \gamma^0 \gamma^s   ( p_{s} + \frac{1}{2 E f}
\{p_{s}, E f\}_- )\}\psi=0.
\label{weylE}
\end{eqnarray}
(We use $s,t,$ or $\sigma, \tau,$ to denote the two compactified dimensions ($x^5$ and $x^6$ for 
the flat and the Einstein index, respectively), $m,n,$ to denote  
the flat experimentally observed  $1+3= d-m$ dimensions, and $a,b,$ and $\alpha, \beta,$ to 
denote all the flat and Einstein indices, respectively.) 
The authors of this work found a way out of the ``Witten's no go theorem'' 
 for a toy model of 
$M^{(1+3)} \times$ a flat finite disk in $(1+5)$-dimensional space~\cite{hnkk06} by postulating a 
particular boundary condition, which allows a spinor to carry  only one handedness 
after the compactification---but many charges (all positive integers). 
Massless spinors then chirally couple to the corresponding background gauge gravitational field, 
which solves equations of motion for a free field, linear in the Riemann curvature, while 
the current through the boundary for the massless and all the massive solutions is equal to zero. 
In ref.~\cite{hnkk06} the boundary condition was written in a covariant way  as 
\begin{eqnarray}
\hat{\cal{R}}\psi|_{\rm wall} &=& 0,\quad 
\hat{\cal{R}} = \frac{1}{2}(1-i n^{(\rho)}{\!\!}_{a}\, n^{(\phi)}{\!\!}_{b}\, 
\gamma^a \gamma^b ),\quad \hat{\cal{R}}^2 = \hat{\cal{R}}
\label{diskboundary}
\end{eqnarray}
with $n^{(\rho)}=(0,0,0,0,\cos \phi, \sin \phi),\; n^{(\phi)}= 
(0,0,0,0,-\sin \phi, \cos \phi)$, which  
are the two unit vectors 
perpendicular and tangential to the boundary of the disk at $\rho_0$, respectively. 
The operator $\hat{\cal{R}}$ is a projector. It can for the above choice of the two vectors 
$n^{(\rho)}$ and $n^{(\phi)}$ be written as  
\begin{eqnarray}
\hat{\cal{R}}   &=& {\stackrel{56}{[-]} }= \frac{1}{2} (1-i\gamma^5\gamma^6).
\label{prodiskboundary}
\end{eqnarray}
In Appendix~\ref{appendixtechnique}  properties of the Clifford algebra 
objects  $\stackrel{ab}{(\pm)}, \stackrel{ab}{[\pm]}$ are presented. 
The boundary condition requires that only  massless states (fulfilling equations 
of motion~(Eq.(\ref{weylE}))) of one (let us say right) handedness 
with respect to the compactified disk  are allowed. Accordingly 
massless states of only one handedness are allowed in $d=(1+3)$ for each charge. 
Since the realistic theory, 
leading to massless states of only one charge is desirable, we search for boundary conditions, 
which would lead to one massless state of one handedness and only one charge.

In this paper: \\
i.  We formulate the  boundary condition allowing one massless state of one handedness 
and only one charge    
and infinitely many massive states with the same charge so that to each mass only 
one state corresponds. \\
ii. We define the operator for momentum $p^a$ so that it becomes 
Hermitean on the vector space of states fulfilling the boundary conditions 
and we comment on the orthogonality relations of these states. \\
iii. We study the properties of states on a curved disc with the boundary.


%

\section{Boundary conditions}
\label{boundaries}

The boundary condition, presented in ref.~\cite{hnm06}, allows 
massless spinors of only one handedness but of all positive 
half integer charges. To resemble with our toy model a "realistic theory" at $d(=1+3)$ 
the massless spinor must manifest in $d=(1+3)$ only one charge. The boundary condition on a 
disc which does lead to massless 
spinors of only one handedness and only one charge and which leads to 
infinite many massive spinors of the same charge is as follows
\begin{eqnarray}
\hat{\cal{R'}}\psi|_{\rm wall} &=& 0,\nonumber\\
\hat{\cal{R'}} &=& \frac{1}{2}( \stackrel{56}{[-]} + 
 e^{i \theta}\; n^{(\phi)a} \,   p_{a} \; n^{(\phi)}{}_{b} \, \gamma^{b} \; \stackrel{56}{[-]} \; 
n^{(\rho)}{}_{c} \, \gamma^{c} \; n^{(\rho)d}           \, p_{d} + h.c. ) 
\label{diskboundarynew}
\end{eqnarray}
with $\stackrel{56}{[-]} =(1-i n^{(\rho)}{\!\!}_{a}\, n^{(\phi)}{\!\!}_{b}\, 
\gamma^a \gamma^b ),$ while $\theta$ is a parameter. 
If taking $n^{(\rho)a}= (0,0,0,0,\cos \phi, \sin \phi)$ and 
$n^{(\phi)a}= (0,0,0,0, -\sin \phi, \cos \phi)$,  $\hat{\cal{R'}}$ in 
Eq.(\ref{diskboundarynew}) simplifies to 
\begin{eqnarray}
\hat{\cal{R'}} = \stackrel{56}{[-]}\; + \cos \theta \; \stackrel{56}{[+]}\; 
(- \frac{i}{\rho} \,\frac{\partial}{\partial \rho}\,\frac{\partial}{\partial \phi} ). 
\label{diskboundarynew1}
\end{eqnarray}
The two conditions, the one of Eq.(\ref{diskboundarynew1}) and that of Eq.(\ref{prodiskboundary}), 
coincide for $\theta = (2k+1)\, \pi/2$, where $k$ is an integer. 
However, while $\hat{\cal{R}}$ is a projector, 
$\hat{\cal{R'}}$ is not. 
Both, $\hat{\cal{R}}$ and $\hat{\cal{R'}}$, are Hermitean.  
Accordingly $\hat{\cal{O}} = I- 2\hat{\cal{R}}$, with $\hat{\cal{O}}^2= I $ and 
$\hat{\cal{O}}^{\dagger}=\hat{\cal{O}}$, is unitary, while $\hat{\cal{O'}} = 
I - 2\hat{\cal{R'}}$ is not unitary. We shall see that when requiring  that 
solutions of equations of motion obey  the condition $\hat{\cal{R'}}^k\psi|_{\rho=\rho_0} = 0, $ 
for $k=1$, the same states are projected out for any $k$. In this sense also $\hat{\cal{O'}}$ 
manifests  
unitarity on the states of solutions (obeying boundary conditions on the boundary, 
where both operators only apply).

\section{Equations of motion and solutions}
\label{equations}

We study two cases: \\
i. We assume that the two dimensional space of coordinates $x^5$ and $x^6$ is 
a Euclidean plane $M^{(2)}$ (with no gravity) so that   
$f^{\sigma}{\!}_{s} = \delta^{\sigma}{\!}_{s},\; \omega_{56 s} =0$,  
with the rotational symmetry around the origin, \\
ii. We assume that space of $x^5$ and $x^6$ is curved with the zweibein 
$f^{\sigma}{\!}_{s}= \delta^{\sigma}{\!}_{s} f(\rho), $ 
with $\rho,$ defined by
%
$x^5= \rho \cos \phi,\quad x^6= \rho \sin \phi,$ 
%
so that the rotational symmetry around the axis 
perpendicular to the plane of 
$x^5$ and $x^6$ is preserved.

If the Euclidean plane is curved on $S^2$ with the radius $\rho_0$ and  
the rotational symmetry around the axis 
perpendicular to the plane of 
$x^5$ and $x^6$, 
then $f(\rho)= 1+ (\frac{\rho}{\rho_0})^2$. From $ds^2= 
e_{s \sigma}e^{s}{\!}_{\tau} dx^{\sigma} dx^{\tau}= f^{-2}(d\rho^{2} + \rho^2 d\phi^{3})$ 
we find $E= f^{-2}(\rho)$.

Wave functions  describing spinors in $(1+5)$-dimensional space demonstrating  
$M^{(1+3)}$ $\times$ a disk  symmetry are 
required to obey Eq.(\ref{weylE}). 

The most general solution for a free particle in $d=(1+5)$ should be written as a superposition
of all  four ($2^{6/2 -1}$) states of a single Weyl representation. The reader can see in 
Appendix \ref{appendixtechnique}   technical details about how to write 
a Weyl representation 
in terms of the Clifford algebra objects after making a choice of the Cartan subalgebra,  
for which we take: $S^{03}, S^{12}, S^{56}$. 
In our technique \cite{holgernorma2002} one spinor representation---the four 
states which are the eigenstates of the chosen Cartan subalgebra---are  
the following four products of projections $\stackrel{ab}{[k]}$ and nilpotents 
$\stackrel{ab}{(k)}$: 
\begin{eqnarray}
\varphi^{1}_{1} &=& \stackrel{56}{(+)} \stackrel{03}{(+i)} \stackrel{12}{(+)}\psi_0,\nonumber\\
\varphi^{1}_{2} &=&\stackrel{56}{(+)}  \stackrel{03}{[-i]} \stackrel{12}{[-]}\psi_0,\nonumber\\
\varphi^{2}_{1} &=&\stackrel{56}{[-]}  \stackrel{03}{[-i]} \stackrel{12}{(+)}\psi_0,\nonumber\\
\varphi^{2}_{2} &=&\stackrel{56}{[-]} \stackrel{03}{(+i)} \stackrel{12}{[-]}\psi_0,
\label{weylrep}
\end{eqnarray}
where  $\psi_0$ is a vacuum state.
If we write the operators of handedness in $d=(1+5)$ as $\Gamma^{(1+5)} = \gamma^0 \gamma^1 
\gamma^2 \gamma^3 \gamma^5 \gamma^6$ ($= 2^3 i S^{03} S^{12} S^{56}$), in $d=(1+3)$ 
as $\Gamma^{(1+3)}= -i\gamma^0\gamma^1\gamma^2\gamma^3 $ ($= 2^2 i S^{03} S^{12}$) 
and in the two dimensional space as $\Gamma^{(2)} = i\gamma^5 \gamma^6$ 
($= 2 S^{56}$), we find that all four states are left handed with respect to 
$\Gamma^{(1+5)}$, with the eigenvalue $-1$, the first two states are right handed and the second two 
 states are left handed with respect to 
$\Gamma^{(2)}$, with  the eigenvalues $1$ and $-1$, respectively, while the first two are 
left handed 
and the second two right handed with respect to $\Gamma^{(1+3)}$ with the eigenvalues $-1$ and $1$, 
respectively. 
Taking into account Eq.(\ref{weylrep}) we may write the most general wave function  
$\psi^{(6)}$ obeying Eq.(\ref{weylE}) in $d=(1+5)$ as
\begin{eqnarray}
\psi^{(6)} = \A \,{\stackrel{56}{(+)}}\,\psi^{(4)}_{(+)} + 
\B \,{\stackrel{56}{[-]}}\, \psi^{(4)}_{[-]}, 
\label{psi6}
\end{eqnarray}
where $\A$ and $\B$ depend on $x^5$ and $x^6$, while $\psi^{(4)}_{(+)}$ 
and $\psi^{(4)}_{[-]}$  determine the spin 
and the coordinate dependent parts of the wave function $\psi^{(6)}$ in $d=(1+3)$ 
\begin{eqnarray}
\psi^{(4)}_{(+)} &=& \alpha_+ \; {\stackrel{03}{(+i)}}\, {\stackrel{12}{(+)}} + 
\beta_+ \; {\stackrel{03}{[-i]}}\, {\stackrel{12}{[-]}}, \nonumber\\ 
\psi^{(4)}_{[-]}&=& \alpha_- \; {\stackrel{03}{[-i]}}\, {\stackrel{12}{(+)}} + 
\beta_- \; {\stackrel{03}{(+i)}}\, {\stackrel{12}{[-]}}. 
\label{psi4}
\end{eqnarray}

Using $\psi^{(6)}$ in Eq.(\ref{weylE}) we recognize the following expressions as the mass terms:  
$\frac{\alpha_+}{\alpha_-} (p^0-p^3) - \frac{\beta_+}{\alpha_-} (p^1-ip^2)= m,$ 
$\frac{\beta_+}{\beta_-} (p^0+p^3) - \frac{\alpha_+}{\beta_-} (p^1+ip^2)= m,$ 
$\frac{\alpha_-}{\alpha_+} (p^0+p^3) + \frac{\beta_-}{\alpha_+} (p^1-ip^2)= m,$
$\frac{\beta_-}{\beta_+} (p^0-p^3) + \frac{\alpha_-}{\beta_+} (p^1-ip^2)= m.$ 
(One can notice that for massless solutions  ($m=0$) the $\psi^{(4)}_{+}$ and $\psi^{(4)}_{-}$ 
decouple.) 
We end up with the equations of motion
 for $\A$ and $\B$ as follow 
\begin{eqnarray}
-2i\,f\, (\frac{\partial}{\partial z} + \frac{\partial \ln \sqrt{Ef}}{\partial z})\, \B  
+ m \;\A =0,\nonumber\\
-2i\,f\, (\frac{\partial}{\partial \bar{z}} + \frac{\partial\ln \sqrt{Ef}}{\partial \bar{z}})\, \A  
+ m \;\B =0,  
\label{equationm56gen}
\end{eqnarray}
where $z: = x^5 + i x^6 = \rho \,e^{i\phi}$, $\bar{z}: = x^5 - i x^6 = \rho \, e^{-i\phi}$ 
and $\frac{\partial}{\partial z}   = \frac{1}{2}\,(\frac{\partial}{\partial x^5} - i 
\frac{\partial}{\partial x^6}) = e^{-i\phi} \;
(\frac{\partial}{\partial \rho} - \frac{i}{\rho}\,\frac{\partial}{\partial \phi} ) $, 
$\frac{\partial}{\partial \bar{z}} = \frac{1}{2}\, (\frac{\partial}{\partial x^5} + i 
\frac{\partial}{\partial x^6}) = e^{ i\phi} \;
(\frac{\partial}{\partial \rho} + \frac{i}{\rho}\,\frac{\partial}{\partial \phi} ) $. 
We can rewrite Eq.(\ref{equationm56gen}) in a more compact form as follows
\begin{eqnarray}
-2i\,f\, (\frac{\partial}{\partial z} (\B \sqrt{Ef})  
+ m \;(\A\sqrt{Ef}) =0,\nonumber\\
-2i\,f\, (\frac{\partial}{\partial \bar{z}} (\A \sqrt{Ef})   
+ m \;(\B\sqrt{Ef}) =0.  
\label{equationm56gen1}
\end{eqnarray}
Having the rotational symmetry around the axis perpendicular on the fifth and the sixth 
dimension we require that $\psi^{(6)}$ is the eigenfunction of the total angular momentum
operator $M^{56}$
\begin{eqnarray}
M^{56}\psi^{(6)}= (n+\frac{1}{2})\,\psi^{(6)}, \quad M^{56} = x^5 p^6-x^6 p^5  + 
S^{56}.
\label{mabx}
\end{eqnarray}
Then  $\A=\A_n(\rho)\; e^{i n \phi} $ and $\B= \B_n(\rho)\; e^{i (n+1) \phi}$.

Spinors which manifest masslessness in $d=(1+3)$ must obey the 
equations (\ref{equationm56gen},\ref{equationm56gen1}) for $m=0$. 
One easily sees that for $m=0$ and any $f$, which determines the curvature in the 
fifth and  the sixth dimension ($f^{\sigma}{\,}_{s}= \delta^{\sigma}_s f(\rho)$, while 
$\omega_{abs}=0$),  we find the solution of Eq.(\ref{equationm56gen1}) with the 
total angular momentum in the fifth and the sixth dimension equal to $n+1/2$ as follows 
\begin{eqnarray}
\label{m0sol}
\psi^{(6)\, n+1/2}_{0}= \alpha_n \; z^n \; \sqrt{f}\;\stackrel{56}{(+)}\;\psi^{(4)}_{(+) 0} + 
                \beta_n  \; \bar{z}^n \; \sqrt{f}\;\stackrel{56}{[-]}\;\psi^{(4)}_{[-]0}.
\end{eqnarray}
There is a solution for any positive integer $n$ and there is obviously no mass protection, 
since the solution for any chosen $n$ is the superposition of the left and the right handed 
components in $d=(1+3).$
Taking into account the boundary condition of Eq.(\ref{diskboundary}) one  sees that  
$\beta_n$ must be zero for all $n$, accordingly $\psi^{(6)n+1/2}_{0}= 
\alpha_n \; z^n \; \sqrt{f} \;\stackrel{56}{(+)}\;\psi^{(4)}_{(+) 0}$ for any $f(\rho)$, which 
means that only solutions of one handedness in $d=(1+3)$ are allowed, assuring the
 mass protection mechanism in $d=(1+3)$. However, all the positive integers $n$ are allowed 
 (we require $n\ge 0$ to ensure the integrability of solutions at the origin). 

Taking into account the boundary condition of Eq.(\ref{diskboundarynew1}) we see that 
$\beta_n$ must still be zero for any $n$, while now the condition  
$\hat{\cal{R'}}\; \psi^{(6)\, n+1/2}_{0}|_{\rho=\rho_0}=0 $ leads to  the condition 
\begin{eqnarray}
\label{Rc}
n (n + \rho \frac{\partial \ln \sqrt{f}}{\partial \rho})|_{\rho =\rho_0}=0.
\end{eqnarray}
In the case that a disc is flat with the boundary at $\rho =\rho_0$ we get 
that only $n=0$ fulfils the boundary condition of Eqs.(\ref{diskboundarynew1},\ref{Rc}). 
If a disk is
curved on a sphere with radius $\rho_0$ and we put a boundary at $\rho=\rho_0$, 
then $f= (1+ (\frac{\rho}{\rho_0})^2)$ and the boundary condition requires 
$n(n + \frac{1}{2})=0$. Again the only solution is $n=0,$ since $n$ is an integer. 
More general $f$ would lead to rational or irrational numbers (and so would 
a boundary at some other $\rho_1 \ne \rho_0$), 
so that we can conclude 
that $n=0$ is the only solution even if the disk is not flat.

Therefore for $m=0$ we get as the only solution for any curvature $f$  ($f^{\sigma}{\!}_{s} = f\,
\delta^{\sigma}{\!}_{s},\; \omega_{56 s} =0,\, s=5,6; \sigma =5,6$,)
\begin{eqnarray}
\psi^{(6)\, 1/2}_{0} &=& a_0 \stackrel{56}{(+)} \psi^{(4)}_{(+) 0}
\label{solmeq0}
\end{eqnarray}
and accordingly this massless state is mass protected and manifests spin $1/2$ (as we shall see)
as the only charge in $d=(1+3)$ (Eq.(\ref{mabx})).

In the  massive case ($m \ne 0$) we get for $f=1$ the following solutions of 
Eq.(\ref{equationm56gen1})  
\begin{eqnarray}
\psi^{(6)\, n+1/2}_{m} &=& {\cal N}_n \; e^{in \phi}\; \{ J_n  \;\stackrel{56}{(+)}\; \psi^{(4)}_{(+) m}
-i e^{i \phi} \; J_{n+1} \;\stackrel{56}{[-]}\; \psi^{(4)}_{[-] m} \}, 
\label{massive0}
\end{eqnarray}
where $J_{n}$ are the Bessel's functions of the first order, which depend  
on $\rho$, while ${\cal N}_n$ determines the normalization~\cite{hnm06}.

If we require that the boundary condition of Eq.(\ref{prodiskboundary}) should be fulfilled, then 
$J_{n+1}|_{\rho=\rho_0} =0$ and the zeros of $J_{n+1}$ determine for each $n$ and each zero of 
$J_n$  a (different) mass $m$.

If we require that the boundary condition of Eq.(\ref{diskboundarynew1}) is fulfilled, then 
only $n=0$ is the solution.  
In this case 
we namely have
$J_{n+1}|_{\rho=\rho_0}= 0$ and $ n \; (\frac{1}{\rho} \; 
\frac{\partial J_{n}}{\partial \rho} )|_{\rho=\rho_0}=0$. It turns out that $n=0$ is the only 
possibility, since   $J_{n+1}|_{\rho=\rho_0}= 0= 
\frac{\partial J_{n}}{\partial \rho}|_{\rho=\rho_0}$ is true only for $n=0$. 
This relation is fulfilled for infinitely many masses $m_{i}\,=\,\alpha_{1i}/\rho_0,\,
i=1,\cdots$, where index $i$ 
determines the successive number of a zero of $J_1$ at $\rho= \rho_0$. 
There are accordingly infinitely many massive solutions, which obey the equations of 
motion (Eq.(\ref{equationm56gen1})) for $f=1$ 
and the boundary condition of Eq.(\ref{diskboundarynew1}), all having eigenvalue of $M^{56}$ 
equal to $1/2$ 
\begin{eqnarray}
\label{massive}
\psi^{(6)\, 1/2}_{m^i}= {\cal N}_i \; (J_{0}(\alpha_{1i} \rho/\rho_0) \;
\stackrel{56}{(+)} \; \psi^{(4)}_{(+) m^i} - iJ_{1}\;
(\alpha_{1i} \rho/\rho_0) \;
\stackrel{56}{[-]} \; \psi^{(4)}_{[-] m^i}).
\end{eqnarray}

For   $f = (1+ (\frac{\rho}{\rho_0})^2)$ the solutions of Eq.(\ref{equationm56gen1}) 
obeying the boundary condition of Eq.(\ref{diskboundarynew1}) can not be found among the 
known functions, but we still know that they have the eigenvalue of $M^{56}$ equal to 
$1/2$ and we also guess that they behave pretty like the two Bessel's functions in 
Eq.(\ref{massive}).

\section{Current through the wall}
\label{current}

The current perpendicular to the wall can be written as
\begin{eqnarray}
n^{(\rho)s} j_s &=&\psi^{\dagger} \gamma^0 \gamma^s n^{(\rho)}_s \psi =  
\psi^{\dagger}\hat{j}_{\perp} \psi, 
\quad \hat{j}_{\perp} = - \gamma^0 \{ e^{-i\phi} \stackrel{56}{(+)} + 
e^{i\phi} \stackrel{56}{(-)}\}. 
\label{current}
\end{eqnarray}
For physically acceptable cases when 
spinors are localized inside the disk 
 the current through the wall must be equal to zero 
\begin{eqnarray}
\{\psi^{\dagger}\hat{j}_{\perp}\psi\}|_{\rho=\rho_0} = 0. 
\label{current}
\end{eqnarray}
One easily checks that---since the current operator $\hat{j}_{\perp}$
changes the handedness of a state---in the massless case (since  massless states 
of only one handedness exist (Eq.(\ref{solmeq0})) and all the massive cases  
(the product of $\A$ and $ \B$ appears  
in the current and $\B$ is zero, in particular in Eq.(\ref{massive}) 
$J_{1i}|_{\rho = \rho_0} =0$)
the current through the wall is for both types of the boundary conditions equal to zero. 


%

\section{Hermiticity of the operators and orthogonality of solutions}
\label{hermiticity}

The operators $ p_s$ (and consequently also 
$(\gamma^s p_s)^2$) are not Hermitean on the space of solutions which have nonzero values 
on the boundary $\rho=\rho_0$, since then $\int d^2x p_s(\psi_{i}{\!}^{\dagger}  \psi_{j})\ne 0$. 
We define therefore a new operator $\hat{p}_s$. We take care of a flat disc with the boundary. 

{\it Statement:} The operators $\hat{p}_s$, 
   \begin{eqnarray}
   \hat{p}_s= i\{ \frac{\partial}{\partial x^s} -  \frac{1}{2} \frac{x^s}{\rho}
   \delta(\rho-\rho_0)\stackrel{56}{[+]}\},  
   \label{ps}
   \end{eqnarray}
are Hermitean on the vector space of solutions presented in Eqs.(\ref{solmeq0},\ref{massive}).

{\it Proof:} Since the expectation value of $\hat{p}_s$ is zero between the 
massless states (Eq.(\ref{solmeq0})) and so is also between the massless and all the 
massive states (Eq.(\ref{massive})), we check the Hermiticity of the operator 
 $(\gamma^s \hat{p}_s)^2 = p_s p^s \;+\;
  \frac{1}{2}\{ (\frac{\partial}{\partial \rho} + \frac{1}{\rho} ) \delta(\rho -\rho_0) 
  + \delta(\rho-\rho_0)
  ( \frac{\partial}{\partial \rho} - \frac{i}{\rho} \frac{\partial}{\partial \phi})
  \},$   
where $p_s p^s= \frac{\partial^2}{\partial \rho^2} + \frac{1}{\rho^2}
\frac{\partial^2}{\partial \phi^2} +  \frac{1}{\rho}
\frac{\partial}{\partial \phi}$ ($ \int d^2x 
 \psi_{i}{\!}^{\dagger}(\gamma^s p_s)^2)  \psi_j = 
\int d^2x \psi_{i}{\!}^{\dagger}  \psi_j m^2  \delta_{ij}$).  
We ought to check  only the $x^5$ and $x^6$ part, with  
the corresponding spin components included. We obtain for the expectation values 
of $(\gamma^s \hat{p}_s)^2 $ between
the massless and a massive state the values:  
$\;\int d^2x \Tr\Sb{56} (J_{0i}\stackrel{56}{(+)})^{\dagger} 
[(\gamma^s \hat{p}_s)^2 
\stackrel{56}{(+)}] \,=\,  \pi \frac{\alpha_{1i}}{\rho_0} (\rho J_{1i})|_{\rho=\rho_0} \,=
\, \int d^2x \Tr\Sb{56} [(\gamma^s \hat{p}_s )^2 J_{0i}\stackrel{56}{(+)}]^{\dagger}  
\stackrel{56}{(+)}\, = \,\pi (-\frac{\alpha_{1i}}{\rho_0} \rho J_{1i})|_{\rho=\rho_0}\, = \,0$,  
due to the properties 
of the Bessel functions $J_{1i} \,= - \frac{\rho_0}{\alpha_{1i}} 
\frac{\partial J_{0i}}{\partial \rho}$ with 
$J_{1i}(\alpha_{1i})\,= 0$ and the property of the delta function $\int_{0}^{\rho_0} g(\rho) 
\frac{\partial \delta(\rho -\rho_0)}{\partial \rho}\, =\,
- \frac{\partial g(\rho)}{\partial \rho}|_{\rho=\rho_0}$, where 
$g(\rho)$ is any smooth function of $\rho$. 
Accordingly the massless state 
is orthogonal to all the massive states. 
Taking into account that for the  Bessel functions 
$\frac{\partial J_{1i}}{\partial \rho} = \frac{\alpha_{1i}}{\rho_0}
J_{0i} + \frac{\rho_0}{\alpha_{1i}} \frac{1}{\rho} \frac{\partial J_{0i}}{\partial \rho}$
one finds that for $i\ne k$ it follows  
\begin{eqnarray}
\label{herorth1}
&&\int d^2x \Tr\Sb{56} 
(J_{0i}\stackrel{56}{(+)} -iJ_{1i} \stackrel{56}{[-]} e^{i \phi})^{\dagger} 
[(\gamma^s \hat{p}_s)^2 (J_{0k}\stackrel{56}{(+)}-i J_{1k} \stackrel{56}{[-]} e^{i \phi})]=\nonumber\\ 
&&2 \pi \{\int^{\rho_0}_0 \rho d\rho [- (m_{k})^2 (J_{0i}J_{0k} + J_{1i}J_{1k})] 
+ \; \frac{1}{2} ( -\rho \frac{\partial J_{0i}}{\partial \rho} J_{0k} + \rho J_{1i} J_{0k} 
\frac{\alpha_{1k}}{\rho_0})|_{\rho = \rho_0} \}= \nonumber\\
&&2\pi(\rho J_{0i}J_{1k} + 
\rho J_{1i}J_{0k})|_{\rho=\rho_0} =0, 
\end{eqnarray} 
 since $ J_{1k} (\alpha_{1k})=0.$ We checked accordingly 
the Hermiticity of the operator $\rho(\gamma^s \hat{p}_s )^2 $ on the vector space of 
the massive states and correspondingly  also the orthogonality of these states.

Let us add  the normalization property of the massive states 
\begin{eqnarray}
\label{orth0}
&&\int d^2x \Tr\Sb{56} 
(J_{0i}\stackrel{56}{(+)} -iJ_{1i} \stackrel{56}{[-]} e^{i \phi})^{\dagger} 
(\gamma^s \hat{p}_s )^2 (J_{0i}\stackrel{56}{(+)}-i J_{1i} \stackrel{56}{[-]} e^{i \phi})=
\nonumber\\
&&- \pi \,(m_{k})^2 (\rho^2 (J_{0i}^2  +  J_{1i}^2))|_{\rho = \rho_0} = 
\pi (\rho^2 J_{0i}^2)|_{\rho=\rho_0},
\end{eqnarray}
since $m_{i}= \frac{\alpha_0i}{\rho_0}$.

We conclude that on the space of  solutions (Eqs.(\ref{solmeq0},\ref{massive})) the operators 
$\hat{p}_s$ (Eq.(\ref{ps})) are Hermitean and the solutions are orthogonal. Since we do not 
know the explicit expressions   for solutions on the curved disc ($f\ne1$), we do not comment 
orthogonality properties of these functions.

%

%

\section{Properties of spinors in $d=(1+3)$}
\label{properties1+3}

To study how do spinors couple to the Kaluza-Klein gauge fields in the case of $M^{(1+5)}$, ``broken'' to 
$M^{(1+3)} \times $ a flat disk with $\rho_0$ and with the involution boundary condition, 
which allows only right handed spinors
at $\rho_0$,
we first look for (background) gauge gravitational fields, which preserve the rotational symmetry 
on the disk. Following ref.\ \cite{hnkk06} we find 
for the background vielbein field  
\begin{eqnarray}
e^a{}_{\alpha} = 
\pmatrix{\delta^{m}{}_{\mu}  & e^{m}{}_{\sigma}=0 \cr
 e^{s}{}_{\mu} & e^s{}_{\sigma} \cr},
f^{\alpha}{}_{a} =
\pmatrix{\delta^{\mu}{}_{m}  & f^{\sigma}{}_{m} \cr
0= f^{\mu}{}_{s} & f^{\sigma}{}_{s} \cr},
\label{f6}
\end{eqnarray}
with $f^{\sigma}{}_{m} = A_{\mu} \delta ^{\mu}{}_{m}
\varepsilon^{\sigma}{}_{\tau} x^{\tau}$
and  the spin connection field 
\begin{eqnarray}
\omega_{st \mu} = - \varepsilon_{st}  A_{\mu},\quad \omega_{sm \mu} = 
-\frac{1}{2} F_{\mu \nu} \delta^{\nu}{}_{m}
\varepsilon_{s \sigma} x^{\sigma}.
\label{omega6}
\end{eqnarray}
 The $U(1)$ gauge field $A_{\mu}$ depends only on $x^{\mu}$.
All the other components of the spin connection fields are zero, 
since for simplicity we allow no gravity in
$(1+3)$ dimensional space.

To determine the current, coupled to the Kaluza-Klein gauge fields $A_{\mu}$, we
analyze the spinor action
\begin{eqnarray}
{\cal S} &=& \int \; d^dx E \bar{\psi}^{(6)} \gamma^a p_{0a} \psi^{(6)} = \int \; 
d^dx  \bar{\psi}^{(6)} \gamma^m \delta^{\mu}{}_{m} p_{\mu} \psi^{(6)} + \nonumber\\
&& \int \; d^dx   \bar{\psi}^{(6)} \gamma^m (-)S^{sm} \omega_{sm \mu} \psi^{(6)}  + 
\int \; d^dx  \bar{\psi}^{(6)} \gamma^s \delta^{\sigma}{}_{s} p_{\sigma} \psi^{(6)} +\nonumber\\
&& \int \; d^dx   \bar{\psi}^{(6)} \gamma^m  \delta^{\mu}{}_{m} A_{\mu} 
(\varepsilon^{\sigma}{}_{\tau} x^{\tau}
 p_{\sigma} + S^{56}) \psi^{(6)}.
\label{spinoractioncurrent}
\end{eqnarray}
 $\psi^{(6)}$ are solutions of the Weyl equation in $d=(1+5)$ .
 $E$ is for $f^{\alpha}{}_{a}$ from  equal to $f^{-2}$. 
The first term on the right hand side  of Eq.(\ref{spinoractioncurrent}) is the kinetic term
(together with the last  term defines  
the  covariant derivative $p_{0 \mu}$ in $d=(1+3)$).  
The second term on the right hand side  contributes nothing when the integration over 
the disk is performed, since it is proportional to $x^{\sigma}$ ($\omega_{sm \mu} = -\frac{1}{2}
F_{\mu \nu} \delta^{\nu}{}_{m} \varepsilon_{s \sigma} x^{\sigma}$).

We end up with 
\begin{eqnarray}
j^{\mu} = \int \; d^2x \bar{\psi}^{(6)} \gamma^m \delta^{\mu}{}_{m} M^{56}  \psi^{(6)}
\label{currentdisk}
\end{eqnarray}
as  the current in $d=(1+3)$.  The charge in $d=(1+3)$ is  proportional to the total 
angular momentum  $M^{56} =L^{56} + S^{56}$ on a disk, which for either massless or 
massive spinors equal to $1/2$.


\section{Conclusions}
\label{discussions}

We presented for a toy model the boundary 
condition which makes massless spinors which in  
$M^{1+5}$ carry nothing but a spin to live in $M^{(1+3)} \times $ a 
disk with a boundary and  manifest in $M^{(1+3)}$---if massless---as a left handed spinor 
(with no right handed partner and accordingly mass protected), which  
carries only one kind of the Kaluza-Klein type of charge and   
chirally couples to the corresponding Kaluza-Klein gauge field (so that  after 
the second quantization procedure a particle and an antiparticle of only 
that particular charge and the opposite one appear, respectively). 

We propose the boundary condition 
\begin{eqnarray}
\{\hat{{\cal O'}} \psi&=& \psi \}|_{\rho=\rho)0}, \quad 
{\cal {\hat O}} = I-2 \hat{{\cal R'}}, \nonumber\\
\hat{{\cal R'}} &=& \stackrel{56}{[-]} + \cos \theta \; \stackrel{56}{[+]}\; 
(- \frac{i}{\rho} \,\frac{\partial}{\partial \rho}\,\frac{\partial}{\partial \phi}), 
\label{discboundarynew2}
\end{eqnarray}
which  is in Eq.(\ref{diskboundarynew}) presented in a Lorentz invariant way, 
with $\theta$ which is an arbitrary parameter $\ne (2k+1)\pi/2,$ and $k$ 
is any integer.  
This boundary condition allows in the massless case only the right  
handed spinor to live  
on the disk and accordingly manifests left handedness in $M^{(1+3)}$.  
The massless and the massive solutions have the eigen value of the total angular momentum 
in the fifth and the sixth dimension equal to  $1/2$, 
which then manifests  as a charge in $d=(1+3)$. The massless solution is mass protected. 
When  a  disk is flat, the massless solution is independent of $x^5$ and $x^6$, while 
the massive solutions are expressible  in terms of the Bessel's functions 
$J_{0}(\alpha^i \rho/\rho_0)$ and $J_{1}(\alpha^i \rho/\rho_0)$, defining masses 
 $m^i= \alpha^i/\rho_0$ through the requirement  that the $i-th$ 
zero of $J_1$ is zero at $\rho=\rho_0$. 

We define a generalized momentum 
  \begin{eqnarray}
  \hat{p}_s=  i\{ \frac{\partial}{\partial x^s} -  \frac{1}{2}\pmatrix{\cos{\phi}\cr
  \sin{\phi} \cr} \delta(\rho-\rho_0) \stackrel{56}{[+]}\},
  \label{psc}
  \end{eqnarray}
which is  Hermitean on the vector space of states obeying equations of motion 
(Eq.(\ref{weylE}))  for $f^{\sigma}{}_{s}= \delta^{\sigma}_{s}$ 
and our boundary condition (Eq.(\ref{discboundarynew2})). Accordingly also 
the operator 
$\gamma^s \hat{p}_s \gamma^t \hat{p}_t$ is Hermitean on the same 
vector space, and the  states 
are accordingly orthogonal, with the eigen 
values of this operator which demonstrate 
the masses of states. 

The negative  $-1/2$ charge states appear only after the second quantization 
procedure in agreement with what we observe.  

If the  disc is curved, so that $f^{\sigma}{}_{s}= \delta^{\sigma}_{s} f$, with 
$f= (1+(\frac{\rho}{\rho_0})^2)$ if it is on $S^2$ with a radius $\rho_0$, the solutions  
obeying Eqs.(\ref{weylE},\ref{discboundarynew2}) have similar properties as for $f=1$, 
but in this case we present only the explicit expression for the massless state, while 
the massive ones stayed to be determined.

\section{Acknowledgement } One of the authors (N.S.M.B.) 
would like to warmly thank Jo\v ze Vrabec 
for  fruitful discussions.

\appendix %

\section{Spinor representation technique in terms of Clifford algebra objects}
\label{appendixtechnique}

We define\cite{holgernorma2002} spinor representations as superposition of 
products of the Clifford algebra objects 
$\gamma^a$ so that they are  
eigen states of the chosen Cartan sub algebra of the Lorentz algebra $SO(d)$, 
determined by the generators 
$S^{ab} = i/4 (\gamma^a \gamma^b - \gamma^b \gamma^a)$.
By introducing the notation
\begin{eqnarray}
\stackrel{ab}{(\pm i)}: &=& \frac{1}{2}(\gamma^a \mp  \gamma^b),  \quad 
\stackrel{ab}{[\pm i]}: = \frac{1}{2}(1 \pm \gamma^a \gamma^b), \;{\rm  for} \; \eta^{aa} \eta^{bb} =-1, \nonumber\\
\stackrel{ab}{(\pm )}: &= &\frac{1}{2}(\gamma^a \pm i \gamma^b),  \quad 
\stackrel{ab}{[\pm ]}: = \frac{1}{2}(1 \pm i\gamma^a \gamma^b), \;{\rm for} \; \eta^{aa} \eta^{bb} =1,
\label{eigensab}
\end{eqnarray}
it can be checked that  the above binomials are really ``eigenvectors''  of  the generators 
$S^{ab}$
\begin{eqnarray}
S^{ab} \stackrel{ab}{(k)}: &=&  \frac{k}{2} \stackrel{ab}{(k)}, \quad 
S^{ab} \stackrel{ab}{[k]}:  =  \frac{k}{2} \stackrel{ab}{[k]}.
\label{eigensabev}
\end{eqnarray}
Accordingly we have
\begin{eqnarray}
\stackrel{03}{(\pm i)}: &=& \frac{1}{2}(\gamma^0 \mp  \gamma^3),  \quad 
\stackrel{03}{[\pm i]}: = \frac{1}{2}(1 \pm \gamma^0 \gamma^3), \nonumber\\
\stackrel{12}{(\pm )}: &= &\frac{1}{2}(\gamma^1 \pm i \gamma^2),  \quad 
\stackrel{12}{[\pm ]}: = \frac{1}{2}(1 \pm i\gamma^1 \gamma^2), \nonumber\\
\stackrel{56}{(\pm )}: &= &\frac{1}{2}(\gamma^5 \pm i \gamma^6),  \quad 
\stackrel{56}{[\pm ]}: = \frac{1}{2}(1 \pm i\gamma^5 \gamma^6), \nonumber\\
\label{eigensab031256}
\end{eqnarray}
with eigenvalues of $S^{03}$ equal to $\pm \frac{i}{2}$ for $\stackrel{03}{(\pm i)}$ and 
$\stackrel{03}{[\pm i]}$, and to $\pm \frac{1}{2}$ for  $\stackrel{12}{(\pm )}$ and 
$\stackrel{12}{[\pm ]}$, as well as for $\stackrel{56}{(\pm )}$ and 
$\stackrel{56}{[\pm ]}$. 

We further find 
\begin{eqnarray}
\gamma^a \stackrel{ab}{(k)}&=&\eta^{aa}\stackrel{ab}{[-k]},\quad 
\gamma^b \stackrel{ab}{(k)}= -ik \stackrel{ab}{[-k]}, \nonumber\\
\gamma^a \stackrel{ab}{[k]}&=& \stackrel{ab}{(-k)},\quad \quad \quad
\gamma^b \stackrel{ab}{[k]}= -ik \eta^{aa} \stackrel{ab}{(-k)}.
\label{graphgammaaction}
\end{eqnarray}

We also find 
\begin{eqnarray}
\stackrel{ab}{(k)}\stackrel{ab}{(k)}= 0, & & \stackrel{ab}{(k)}\stackrel{ab}{(-k)}
= \eta^{aa}  \stackrel{ab}{[k]}, \quad 
\stackrel{ab}{[k]}\stackrel{ab}{[k]} =  \stackrel{ab}{[k]}, \;\;\quad \quad
\stackrel{ab}{[k]}\stackrel{ab}{[-k]}= 0, 
 \nonumber\\
\stackrel{ab}{(k)}\stackrel{ab}{[k]} = 0, & &  \stackrel{ab}{[k]}\stackrel{ab}{(k)}
=  \stackrel{ab}{(k)}, \quad \quad \quad
\stackrel{ab}{(k)}\stackrel{ab}{[-k]} =  \stackrel{ab}{(k)},
\quad \quad \stackrel{ab}{[k]}\stackrel{ab}{(-k)} =0.
\label{graphbinoms}
\end{eqnarray}

To represent one Weyl spinor in $d=(1+5)$, one must make a choice of the
operators belonging to the Cartan sub algebra of $3$ elements of the group $SO(1,5)$ 
\begin{eqnarray}
S^{03}, S^{12}, S^{56}.
\label{cartan}
\end{eqnarray}
Any eigenstate of the Cartan sub algebra (Eq.(\ref{cartan})) must be a product of 
three binomials, each of which is an eigenstate of one of the three elements. 
A left handed spinor ($\Gamma^{(1+5)} = -1$) representation with $2^{6/2-1}$ basic states 
is presented in Eq.(\ref{weylrep}). 
For example, the state  $\stackrel{03}{(+ i)}\stackrel{12}{(+)}
\stackrel{56}{(+)}\psi_0,$ where $\psi_0$ is a vacuum state (any, which is not annihilated 
by the operator in front of the state) has the eigenvalues  of $ S^{03}, S^{12} $ and $S^{56}$ equal 
to $\frac{i}{2}$, $ \frac{1}{2}$ and $\frac{1}{2}$, correspondingly. All the other states
of one representation of $SO(1,5)$ follow from this one by just the application of all possible 
$S^(ab)$, which do not belong to the Cartan subalgebra.

\end{document}